\documentclass[structabstract]{aa}  
\usepackage{graphicx}
\usepackage{txfonts}
\begin{document}
   \title{Ultraviolet and extreme-ultraviolet line ratio diagnostics for \ion{O}{iv}}

   \author{F. P. Keenan, P. J. Crockett, K. M. Aggarwal, D. B. Jess and M. Mathioudakis
          }

   \institute{Astrophysics Research Centre, School of 
Mathematics and Physics, Queen's University Belfast, Belfast BT7 1NN, Northern Ireland, UK             \\
              \email{F.Keenan@qub.ac.uk}
                      }

   \date{Received ; accepted }

 
  \abstract
   {}
   {We generate theoretical ultraviolet and extreme-ultraviolet 
emission line ratios for \ion{O}{iv} and show 
their strong versatility as electron temperature and
density diagnostics for astrophysical plasmas.}
   {Recent fully relativistic calculations of radiative rates and electron impact
excitation cross sections for \ion{O}{iv}, supplemented with earlier
data for A-values and proton excitation rates, are used to derive theoretical
\ion{O}{iv}
line intensity ratios for a wide range of electron temperatures and densities. }
   {Diagnostic line ratios involving ultraviolet or 
extreme-ultraviolet transitions in \ion{O}{iv} are presented,
that are 
applicable to a wide variety of astrophysical plasmas ranging from
low density gaseous nebulae to the densest solar and stellar flares. Comparisons with observational
data, where available, show good agreement between theory and experiment, providing support for the
accuracy of the diagnostics. However, diagnostics are also presented involving lines that
are blended in existing astronomical 
spectra, in the hope this might encourage 
further observational studies at higher spectral resolution.}
   {}

   \titlerunning{Line ratio diagnostics for O IV}
   \maketitle
%

\section{Introduction}

Ultraviolet and extreme-ultraviolet emission lines arising from  
transitions in B-like \ion{O}{iv} are detected from a wide 
variety of astronomical sources, ranging from
the Sun (Sandlin et al. 1986) to other stars (Christian et al. 2004),
gaseous nebulae (Feibelman 1997) and supernova remnants
(Blair et al. 1991). 
The diagnostic potential
of these lines to provide electron temperature (T$_{e}$) and density (N$_{e}$)
diagnostics for the emitting plasma was first shown by Flower \&\ Nussbaumer (1975),
who also calculated radiative rates plus electron and proton
impact excitation cross sections for the ion.
Since then, 
many authors have generated atomic data for \ion{O}{iv} 
that have subsequently been used to derive theoretical 
diagnostic line ratios (see Tayal 2006 and references therein).

Very recently, Aggarwal \& Keenan (2008) have employed the fully relativistic
{\sc grasp} and Dirac {\sc rmatrx} codes to calculate radiative rates and
electron impact excitation cross sections, respectively, 
for all transitions among the energetically lowest 75 fine-structure levels of \ion{O}{iv}. 
These results are the most extensive currently available for 
\ion{O}{iv}, and also should be the most reliable, at least for the 
excitation cross sections, as discussed in detail by Aggarwal \& Keenan.
In this paper we use these data, supplemented with
previous highly accurate calculations for radiative rates and 
proton excitation cross sections, 
to derive theoretical \ion{O}{iv} ultraviolet and extreme-ultraviolet
emission line ratios applicable to a wide
range of astrophysical plasmas. We demonstrate the versatility of 
\ion{O}{iv} plasma diagnostics, which can provide temperature and density 
estimates for a wide variety of astronomical sources
ranging from
low density gaseous nebulae up to the densest solar and stellar flares.


\section{Atomic data and theoretical line ratios}

The model ion for \ion{O}{iv} consisted of the 75 
fine-structure levels arising from the 
2s$^{2}$2p, 2s2p$^{2}$, 2p$^{3}$, 2s$^{2}$3$\ell$ ($\ell$ = s, p, d),
2s2p3$\ell$ ($\ell$ = s, p, d) and 2s$^{2}$4$\ell$ ($\ell$ = s, p, d, f) configurations.
Energies for all these levels were obtained from the compilation of
experimental values
by the National Institute of Standards and Technology, which may be found at their website
http://physics.nist.gov/PhysRefData/. Test calculations including additional
levels, such as those arising from the 2s$^{2}$5$\ell$ and 2s2p4$\ell$ configurations,
where found to have a negligible effect 
on the theoretical line ratios considered in the present paper.

Einstein A-coefficients for transitions in \ion{O}{iv} 
were obtained from the following sources: (i)
Galav\'is et al. (1998) for the forbidden  2s$^{2}$2p $^{2}$P$_{1/2}$--2s$^{2}$2p 
$^{2}$P$_{3/2}$ transition; (ii) 
Corr\'eg\'e \& Hibbert (2002) for the 
2s$^{2}$2p $^{2}$P$_{J}$--2s2p$^{2}$ 
$^{4}$P$_{J^{\prime}}$
intercombination lines; (iii)
Corr\'eg\'e \& Hibbert (2004) for allowed and 
intercombination lines among the 
2s$^{2}$2p, 2s2p$^{2}$, 2p$^{3}$ and 2s$^{2}$3$\ell$ ($\ell$ = s, p, d)
levels; (iv) Aggarwal \& Keenan (2008) for all remaining transitions.
For electron impact excitation rates, we have adopted the
results of 
Aggarwal \& Keenan, which contain several improvements
over previous calculations
for \ion{O}{iv} undertaken with the {\sc rmatrx} code (Zhang et al. 1994; Tayal 2006), 
including the greatest number of levels and the largest 
range of partial waves.
They are hence probably the most accurate currently available for this ion, as discussed
in detail by Aggarwal \& Keenan.
However, there is also scope for improvement, mainly due to the fact
that the wavefunctions adopted in the calculations of Aggarwal \& Keenan are not as 
accurate as those of some other workers, such as Tachiev \& Froese-Fischer (2000) and 
Corr\'eg\'e \& Hibbert (2002). Indeed, this is why we have adopted the A-values of
Corr\'eg\'e \& Hibbert (2002, 2004) where possible, 
as their results have an estimated uncertainty of only $\pm$5\%, 
compared to $\pm$20\% for those of Aggarwal \& Keenan.
The limitations in the wavefunctions of Aggarwal \& Keenan
may directly affect the subsequent determination of excitation rates, both
for allowed and forbidden transitions. This is because for the
weak allowed transitions, the A-values of Aggarwal \&
Keenan differ by up to 50\%\ with those of Tachiev \& Froese-Fischer (2000)
and Corr\'eg\'e \& Hibbert (2004), while some of the energy levels show
discrepancies of up to 8\%\ with the experimental values (see Tables 1 and 2 of Aggarwal \&
Keenan). Nevertheless, the estimated accuracy of the Aggarwal \& Keenan excitation rate
data is $\pm$20\%\ for
 a majority of transitions, and 
should be the most reliable currently available for \ion{O}{iv}.

As noted by,
for example, Seaton (1964), excitation by protons will  be important
for the 2s$^{2}$2p $^{2}$P$_{1/2}$--2s$^{2}$2p $^{2}$P$_{3/2}$ transition, and in the 
present analysis we have used the theoretical results of Foster et al. 
(1996). However, Flower \&\ Nussbaumer (1975) have pointed 
out that proton excitation should also  
be included for the 
2s2p$^{2}$ $^{4}$P$_{J}$--2s2p$^{2}$ $^{4}$P$_{J^{\prime}}$ transitions in B-like ions, 
and for these
rates we have adopted the calculations of
Foster et al. (1997). Both the Foster et al. (1996) and Foster et al. (1997)
data are estimated to be accurate to $\pm$10\%.

Using the above atomic data in conjunction
with a recently updated version of the statistical equilibrium
code of Dufton (1977), relative \ion{O}{iv} 
level populations and hence emission line strengths were calculated
for two grids of 
electron temperature (T$_{e}$) and density (N$_{e}$) values.
The first grid (for T$_{e}$ = 10000, 15000, 20000 and 30000\,K; N$_{e}$ = 10$^{2}$--10$^{6}$\,cm$^{-3}$ in steps of 0.1 dex)
is appropriate to nebular plasmas, while the second
(T$_{e}$ = 10$^{4.8}$--10$^{5.6}$\,K in steps of 0.1 dex; 
N$_{e}$ = 10$^{8}$--10$^{14}$\,cm$^{-3}$
in steps of 0.1 dex) is for 
solar/stellar plasma conditions. In particular, the adopted temperature range
for the latter covers that over which \ion{O}{iv} 
has a fractional abundance in ionization equilibrium of 
N(\ion{O}{iv}/N(O) $\geq$ 0.006 (Bryans et al. 2008), and hence should be
appropriate to most coronal-type plasmas.
Our results are too extensive to reproduce here, as with 75 fine-structure levels in our 
calculations we have intensities for 2775 transitions at each of the 
530 possible (T$_{e}$, N$_{e}$) combinations considered. 
However, results 
involving any line pair, in either photon or energy units, are freely available from
one of the authors (FPK) by email on request.
 Given errors in the adopted atomic data of between $\pm$5\%\ and $\pm$20\%\ (see 
above), we would expect our theoretical ratios to be accurate to better than $\pm$15\%.

\section{Results and discussion}

The \ion{O}{iv} intercombination multiplet around 1400\,\AA\ provides excellent
electron density diagnostics for both nebular and solar plasmas, as shown in 
Figs. 1--4, where we plot 
the emission line ratios
R$_{1}$ = I(1407.3\,\AA)/I(1401.1\,\AA), R$_{2}$ = I(1404.7\,\AA)/I(1401.1\,\AA)
and R$_{3}$ = I(1407.3\,\AA)/I(1404.7\,\AA) as a function of T$_{e}$ and N$_{e}$.
The transitions corresponding to these lines are listed in Table 1, as are those for
the other \ion{O}{iv} features discussed in this paper.
We note that the ratios R$_{4}$ = I(1399.7\,\AA)/I(1401.1\,\AA) and R$_{5}$ =
I(1397.2\,\AA)/I(1401.1\,\AA) have the same T$_{e}$ and N$_{e}$
dependence as R$_{1}$ and R$_{2}$, respectively (due to common upper levels), but
with R$_{4}$ = 1.02$\times$R$_{1}$ and R$_{5}$ = 0.130$\times$R$_{2}$.

   \begin{figure}
   \centering
   \includegraphics[width=9cm]{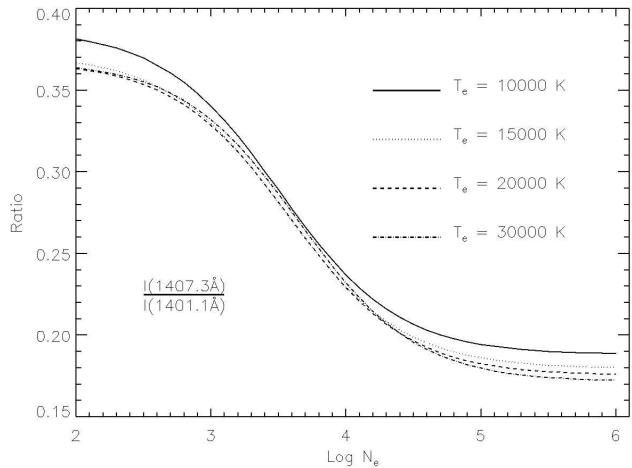}
      \caption{The I(1407.3\,\AA)/I(1401.1\,\AA) line intensity ratio in \ion{O}{iv},
where I is in energy units, plotted as a function of logarithmic electron density
(N$_{e}$ in cm$^{-3}$) at electron temperatures
of T$_{e}$ = 10000, 15000, 20000 and 30000\,K.
              }
         \label{}
   \end{figure}

   \begin{figure}
   \centering
   \includegraphics[width=9cm]{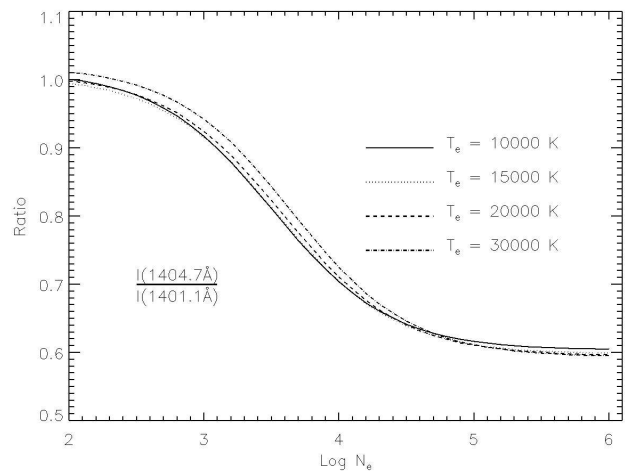}
      \caption{Same as Fig. 1, but for the I(1404.7\,\AA)/I(1401.1\,\AA)
ratio.
              }
         \label{}
   \end{figure}

   \begin{figure}
   \centering
   \includegraphics[width=9cm]{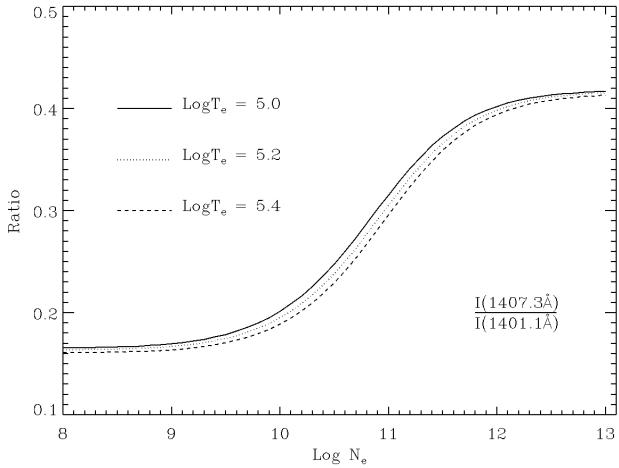}
      \caption{The I(1407.3\,\AA)/I(1401.1\,\AA) line intensity ratio in \ion{O}{iv},
where I is in energy units, plotted as a function of logarithmic electron density
(N$_{e}$ in cm$^{-3}$) at the electron temperature of 
maximum \ion{O}{iv} fractional abundance in ionization equilibrium, T$_{e}$ = 10$^{5.2}$\,K
(Bryans et al. 2008),
plus $\pm$0.2 dex about this value. 
              }
         \label{}
   \end{figure}

   \begin{figure}
   \centering
   \includegraphics[width=9cm]{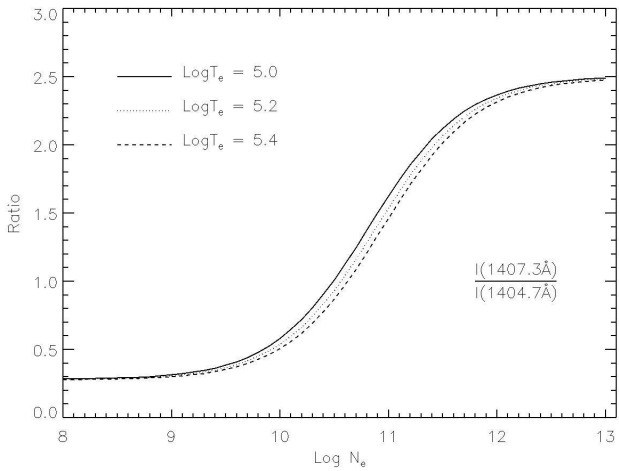}
      \caption{Same as Fig. 3, but for the I(1407.3\,\AA)/I(1404.7\,\AA) ratio.
              }
         \label{}
   \end{figure}

\begin{table}
\caption{Summary of \ion{O}{iv} transitions.}
\label{}
\centering     
\begin{tabular}{l l}
\hline\hline
Wavelength (\AA)      &  Transition 
\\
\hline
554.51  & 2s$^{2}$2p $^{2}$P$_{3/2}$--2s2p$^{2}$ $^{2}$P$_{3/2}$     
\\
625.85  & 2s2p$^{2}$ $^{4}$P$_{5/2}$--2p$^{3}$ $^{4}$S$_{3/2}$     
\\
779.91  & 2s2p$^{2}$ $^{2}$D$_{5/2}$--2p$^{3}$ $^{2}$D$_{5/2}$     
\\
787.71  & 2s$^{2}$2p $^{2}$P$_{1/2}$--2s2p$^{2}$ $^{2}$D$_{3/2}$     
\\
790.20  & 2s$^{2}$2p $^{2}$P$_{3/2}$--2s2p$^{2}$ $^{2}$D$_{5/2}$     
\\
1343.5  & 2s2p$^{2}$ $^{2}$P$_{3/2}$--2p$^{3}$ $^{2}$D$_{5/2}$     
\\
1397.2  & 2s$^{2}$2p $^{2}$P$_{1/2}$--2s2p$^{2}$ $^{4}$P$_{3/2}$     
\\
1399.7  & 2s$^{2}$2p $^{2}$P$_{1/2}$--2s2p$^{2}$ $^{4}$P$_{1/2}$     
\\
1401.1  & 2s$^{2}$2p $^{2}$P$_{3/2}$--2s2p$^{2}$ $^{4}$P$_{5/2}$     
\\
1404.7  & 2s$^{2}$2p $^{2}$P$_{3/2}$--2s2p$^{2}$ $^{4}$P$_{3/2}$     
\\
1407.3  & 2s$^{2}$2p $^{2}$P$_{3/2}$--2s2p$^{2}$ $^{4}$P$_{1/2}$     
\\
\hline
\end{tabular}
\end{table}

The above theoretical ratios are in good agreement 
with observations for gaseous nebulae, such as those by Keenan et al. (1993).
For example, for the planetary nebula NGC\,7662, Keenan et al. measured
R$_{1}$ = 0.30, R$_{2}$ = 0.69 and R$_{4}$ = 0.29,
implying electron densities of 
log N$_{e}$ = 3.5, 4.0 and 3.5, respectively, from Figs. 1 and 2. These values are
consistent, and also in agreement with the electron densities derived for NGC\,7662
from line ratios in species with similar ionization potentials and hence spatial distributions
to \ion{O}{iv}, such as I(4711\,\AA)/I(4740\,\AA) in \ion{Ar}{iv} and 
I(2424\,\AA)/I(2422\,\AA) in \ion{Ne}{iv}, both of which indicate
log N$_{e}$ = 3.5 (Keenan et al. 1997, 1998). 
In addition, we note that Keenan et al. (2002) have measured \ion{O}{iv}
line ratios for the symbiotic star RR\,Tel. The diagnostic ratios for other species
in the RR\,Tel spectrum, ranging from \ion{Al}{ii} 
to \ion{O}{v}, indicate log N$_{e}$
$\simeq$ 5--8 (see Keenan et al. 2002 and references therein), 
over which density interval the theoretical values of 
R$_{1}$ through R$_{5}$ are predicted to be effectively constant (see Figs. 1 and 2).
Once again, there is excellent agreement 
between theory and observation, with 
measured and predicted values of (theory in brackets) R$_{1}$ = 0.16$\pm$0.02 (0.18),
R$_{2}$ = 0.54$\pm$0.05 (0.60), R$_{4}$ = 0.17$\pm$0.02 (0.18) and 
R$_{5}$ = 0.077$\pm$0.008 (0.077).

In existing solar spectra, such as those from the HRTS and SOHO/SUMER
instruments, the \ion{O}{iv} 1404.7\,\AA\ line is significantly blended
with \ion{S}{iv} (Brage et al. 1996; Keenan et al. 2002), although we note that
the blending is negligible under nebular plasma conditions (Keenan et al. 2002). 
Additionally, the 1397.2\,\AA\ feature is very weak in solar spectra, and possibly blended,
and hence its intensity should usually be considered an upper limit (Brage et al. 1996).
However, the remaining 
useable density diagnostic line ratios R$_{1}$ and R$_{4}$ do provide
consistent derivations of N$_{e}$. For example, for Active Region B at +2$^{\prime\prime}$ 
relative to the limb, Feldman \& Doschek (1978)
measured R$_{1}$ = 0.20 and R$_{4}$ = 0.19 from Skylab/S082A spectra, which both imply
log N$_{e}$ = 10.1 from Fig. 3, in good agreement with the value of log N$_{e}$ = 10.2 found from
line ratios in \ion{N}{iv}, which is formed at the same temperature as \ion{O}{iv}
(Keenan et al. 1994).

The intercombination line ratios in \ion{O}{iv}
only provide useful electron density diagnostics 
for values of N$_{e}$ up to about 10$^{12}$\,cm$^{-3}$ (see Figs. 3 and 4).
However, Kastner \& Bhatia (1984) have pointed out that the allowed lines of \ion{O}{iv}
arising from  2s2p$^{2}$ $^{2}$P--2p$^{3}$ $^{2}$D transitions
lie only $\sim$\,60\,\AA\ from the 
intercombination multiplet, and their intensity
ratios are very sensitive to the electron density for N$_{e}$ $>$ 10$^{11}$\,cm$^{-3}$.
They hence should provide good 
N$_{e}$--diagnostics for very high density astronomical plasmas such as solar
and stellar flares.
In Fig. 5 we plot the ratio R$_{6}$ = I(1343.5\,\AA)/I(1407.3\,\AA)
as a function of both T$_{e}$ and N$_{e}$, as the 1343.5\,\AA\ transition is the only
line in the 2s2p$^{2}$ $^{2}$P--2p$^{3}$ $^{2}$D 
multiplet which is unblended in existing solar spectra (Cook et al. 1994). 
The measured values of R$_{6}$ = 0.89 and 0.90 for the flares of 1973 August 9 and 1973 September
7, from Skylab/S082B spectra (Cook et al. 1994), both indicate log N$_{e}$ = 12.5 at the 
temperature of maximum fractional abundance for \ion{O}{iv} in ionization
equilibrium, T$_{e}$ = 10$^{5.2}$\,K (Bryans et al. 2008).
These are consistent with the values of log N$_{e}$ $\simeq$ 12.3 determined for 
high density flares using extreme-ultraviolet lines of \ion{O}{v}, formed at a similar
temperature to \ion{O}{iv} (Keenan et al. 1991).

   \begin{figure}
   \centering
   \includegraphics[width=9cm]{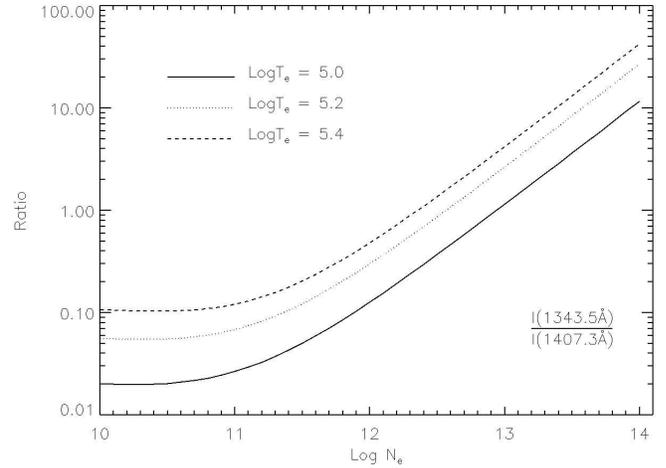}
      \caption{Same as Fig. 3, but for the I(1343.5\,\AA)/I(1407.3\,\AA) ratio.
              }
         \label{}
   \end{figure}

For temperature diagnostics, Flower \& Nussbaumer (1975) have shown that the intensity ratio of lines within the 2s$^{2}$2p $^{2}$P--2s2p$^{2}$ $^{2}$D multiplet at $\sim$\,790\,\AA\ to those in 2s$^{2}$2p $^{2}$P--2s2p$^{2}$ $^{2}$P  at $\sim$\,554\,\AA\ allows T$_{e}$ to be estimated for
the \ion{O}{iv} emitting region of a plasma. This is shown in Fig. 6, where we plot
the R$_{7}$ = I(790.20\,\AA)/I(554.51\,\AA) ratio as a function of T$_{e}$ and N$_{e}$.
Similarly, Curdt et al. (1997) have noted that the R$_{8}$ =  I(779.91\,\AA)/I(787.71\,\AA)
ratio is a T$_{e}$--diagnostic, and this is plotted in Fig. 7. 
An inspection of the two figures reveals that R$_{7}$ is in principle
a better temperature
diagnostic than R$_{8}$, as it is less sensitive to the adopted
value of N$_{e}$. However, the emission lines in R$_{8}$ are much closer in
wavelength and hence the ratio is more likely to be reliably measured.
Indeed, Curdt et al. have determined R$_{8}$ = 0.033 from a SOHO/SUMER
spectrum of the solar disk, which indicates T$_{e}$ $\simeq$ 10$^{5.1}$--10$^{5.3}$\,K 
from Fig. 7 (the exact value depending on the adopted density), in good 
agreement with the temperature of maximum fractional abundance in ionization
equilibrium for \ion{O}{iv}, T$_{e}$ = 10$^{5.2}$\,K (Bryans et al. 2008). Unfortunately, to 
our knowledge, the R$_{7}$ ratio
has not been measured due to the low spectral resolution of existing solar observations
spanning the 554--790\,\AA\ wavelength range. However, O'Shea et al. (1996) have determined
values of the multiplet intensity ratio I(790\,\AA)/I(554\,\AA) for several solar features
from spectra obtained with the S-055 spectrometer on Skylab. These measurements lie in the 
range 0.47--0.56, and imply temperatures within $\sim$\,0.2 dex of that of maximum 
fractional abundance for \ion{O}{iv}.

   \begin{figure}
   \centering
   \includegraphics[width=9cm]{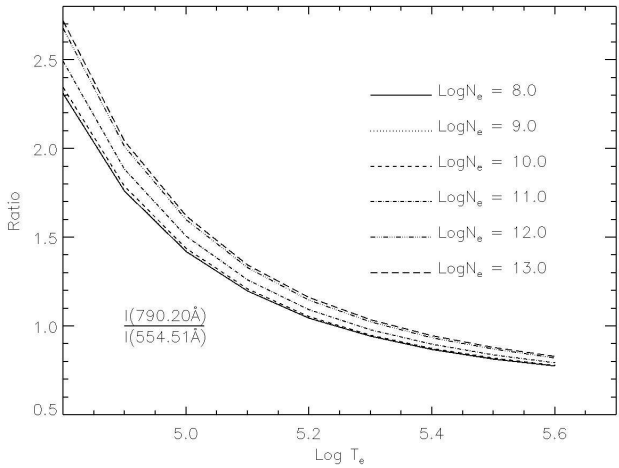}
      \caption{The I(790.20\,\AA)/I(554.51\,\AA) line intensity ratio in \ion{O}{iv}, where I
is in energy units, plotted as a function of logarithmic electron temperature (T$_{e}$
in K) at logarithmic electron densities (N$_{e}$ in cm$^{-3}$) of log N$_{e}$ = 8--13. 
              }
         \label{}
   \end{figure}

   \begin{figure}
   \centering
   \includegraphics[width=9cm]{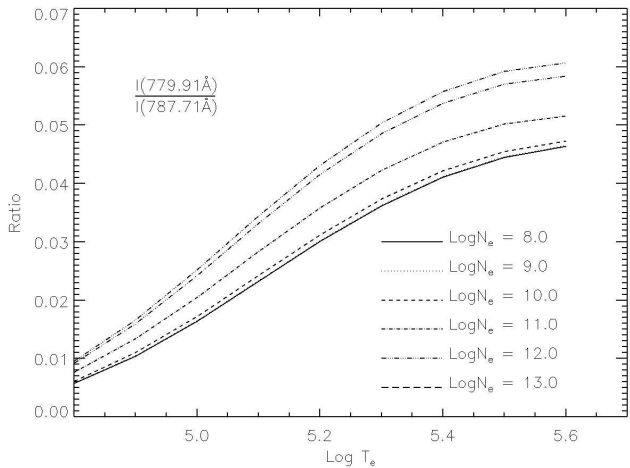}
      \caption{Same as Fig. 6, but for the I(779.91\,\AA)/I(787.71\,\AA) ratio.
              }
         \label{}
   \end{figure}

Under solar plasma conditions, \ion{O}{iv} extreme-ultraviolet line ratios
containing a component of the 2s2p$^{2}$ $^{4}$P--2p$^{3}$ $^{4}$S
multiplet at $\sim$\,625\,\AA\
are strongly density sensitive (O'Shea et al. 1996). Unfortunately, in existing
solar spectra the transitions --- which lie at wavelengths of 
624.62, 625.13 and 625.85\,\AA\ --- are all blended with the strong \ion{Mg}{x}
resonance line at 624.94\,\AA. However, if the \ion{O}{iv}
features could be resolved, then in conjunction with other relatively nearby \ion{O}{iv} lines
they would allow the temperature {\em and} density of the \ion{O}{iv} emitting region of a 
plasma 
to be simultaneously determined via ratio -- ratio 
diagrams, as illustrated in Fig. 8.

\begin{figure}
 \centering
\includegraphics[width=9cm]{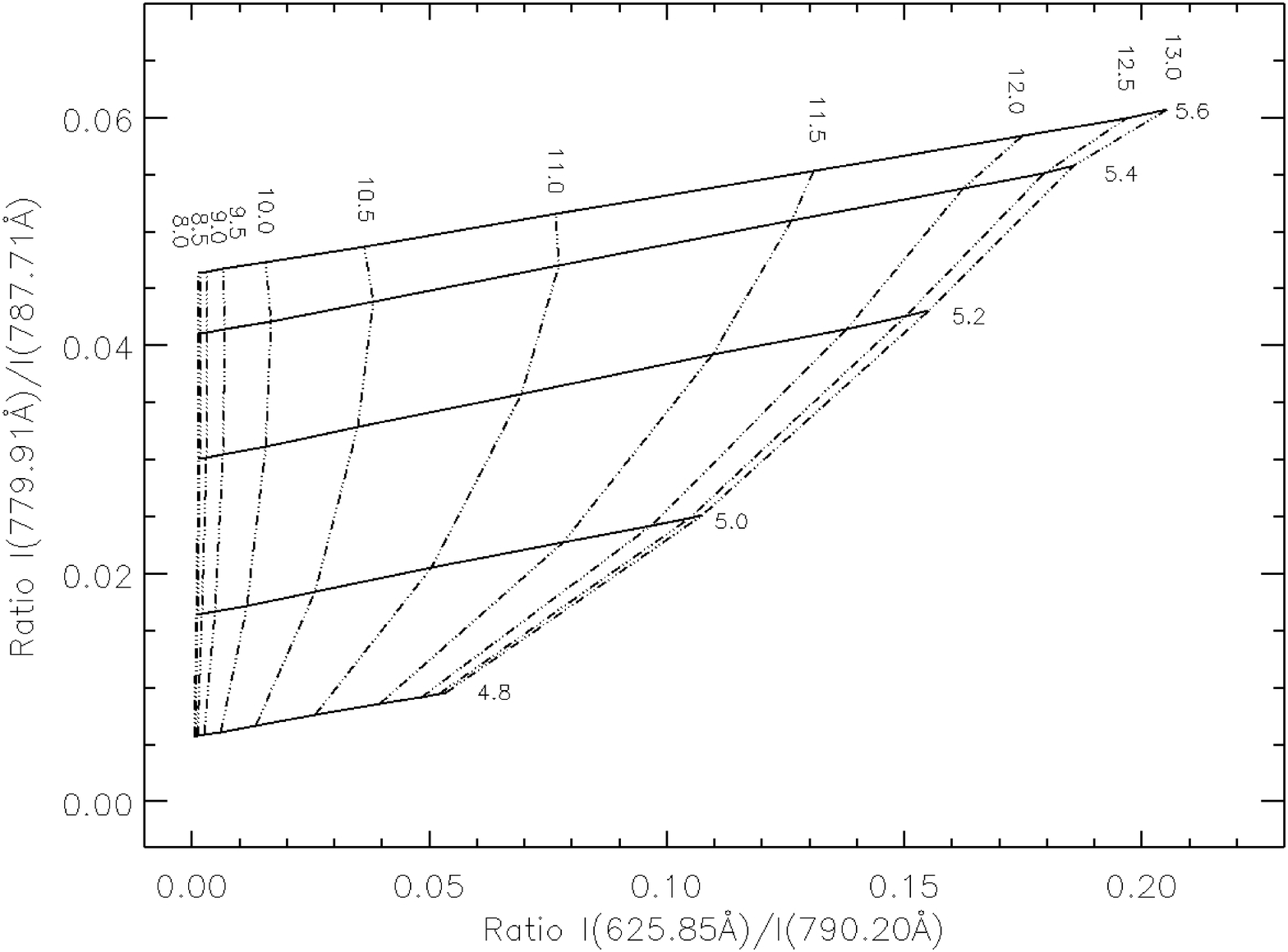}
  \caption{Plot of the \ion{O}{iv} line intensity ratio I(779.91\,\AA)/I(787.71\,\AA)
against I(625.85\,\AA)/I(790.20\,\AA), where I is in energy 
units, for logarithmic electron temperatures (T$_{e}$ in K) of 
log T$_{e}$ = 4.8--5.6, and logarithmic electron densities 
(N$_{e}$ in cm$^{-3}$) of log N$_{e}$ = 8--13.
         }
   \label{}
  \end{figure}

In summary, we see that ultraviolet and extreme-ultraviolet
emission lines of \ion{O}{iv}
provide a diverse portfolio
of T$_{e}$ and N$_{e}$ diagnostics, applicable to a wide variety of
astronomical sources 
ranging from gaseous nebulae to high electron density stellar flares.
The present line ratio calculations, which include the 
most up-to-date atomic physics data, show good agreement
with observations where available, hence providing support for their accuracy.
However, higher spectral resolution observations of several 
\ion{O}{iv} lines, such as the components of the 
2s2p$^{2}$ $^{4}$P--2p$^{3}$ $^{4}$S
multiplet at $\sim$\,625\,\AA, would be very useful to allow their intensities
to be reliably measured and employed as diagnostics.

\begin{acknowledgements}

This work has been financed by the Science and Technology Facilities Council and
Engineering and Physical Sciences
Research Council of the United Kingdom, while
P.J.C. and D.B.J. are 
grateful to the Department of Education and Learning
(Northern Ireland) for the award of studentships.
F.P.K. is grateful to AWE Aldermaston for the award of a William Penney
Fellowship. 

\end{acknowledgements}

\end{document}